\documentclass[aps,prl,reprint]{revtex4-2}

\usepackage[utf8]{inputenc}
\usepackage{amsthm}
\usepackage{amsmath}
\usepackage{color}
\usepackage{hyperref}
\usepackage{graphicx,xcolor}
\usepackage{amssymb}
\usepackage{bbold}

\newcommand{\ii}{\mathrm{i}} 
\newcommand{\eul}{\mathrm{e}} 
\newcommand{\diff}{\mathrm{d}} 
\newcommand{\id}{\mathbb{1}} 

\newcommand{\bvec}[1]{\boldsymbol{#1}} 

\newcommand{\ket}[1]{|#1\rangle} 
\usepackage{tikz}
\usetikzlibrary{quotes}
\usetikzlibrary{arrows}
\usetikzlibrary{fadings}
\tikzfading[name=fade right,
  left color=transparent!0, right color=transparent!100]
\tikzfading[name=fade left,
  left color=transparent!100, right color=transparent!0]

\tikzset{every edge quotes/.style =
          { fill = white,
            sloped,
            font=\footnotesize,
  }}
\tikzstyle{tensor}=[rectangle,draw=blue!50,fill=blue!10,thick]
\tikzstyle{tensor2}=[rectangle,draw=red!50,fill=red!10,thick]
\tikzstyle{vector}=[rectangle,draw=green!80,fill=green!10,thick,rounded corners=.15cm]
\definecolor{green}{HTML}{e88f89}
\definecolor{orange}{HTML}{208c46}
\definecolor{purple}{HTML}{e88f89}
\definecolor{grey}{rgb}{0.75,0.75,0.75}

\newcommand{\braket}[1]{\langle#1\rangle} 
\newcommand{\ketbra}[2]{|#1\rangle\!\langle #2|} 
\newcommand{\tr}{\mathrm{tr}} 

\newcommand{\emphquote}[1]{‘#1’} 

\begin{document}

\title{Open Quantum System Dynamics from Infinite Tensor Network Contraction}
\author{Valentin Link}
\affiliation{Institut f{\"u}r Theoretische Physik, Technische Universit{\"a}t Dresden, 
D-01062, Dresden, Germany}

\author{Hong-Hao Tu}
\affiliation{Institut f{\"u}r Theoretische Physik, Technische Universit{\"a}t Dresden, 
D-01062, Dresden, Germany}

\author{Walter T. Strunz}
\affiliation{Institut f{\"u}r Theoretische Physik, Technische Universit{\"a}t Dresden, 
D-01062, Dresden, Germany}
\email{walter.strunz@tu-dresden.de}

\date{\today}

\begin{abstract}
Approaching the long-time dynamics of non-Markovian open quantum systems presents a challenging task if the bath is strongly coupled. Recent proposals address this problem through a representation of the so-called process tensor in terms of a tensor network. We show that for Gaussian environments highly efficient contraction to matrix product operator (MPO) form can be achieved with infinite MPO evolution methods, leading to significant computational speed-up over existing proposals. The result structurally resembles open system evolution with carefully designed auxiliary degrees of freedom, as in hierarchical or pseudomode methods. Here, however, these degrees of freedom are generated automatically by the MPO evolution algorithm. Moreover, the semi-group form of the resulting propagator enables us to explore steady-state physics, such as phase transitions.

\end{abstract}\maketitle

\paragraph{Introduction} 

Dissipative effects are crucial to our understanding of real world quantum mechanical systems and feature a variety of relevant physical phenomena absent in purely unitary settings. 
In many realistic and particularly interesting setups, the time scales of system and environment do not separate, leading to a buildup of strong correlations with the environment \cite{deVega2017Jan,Thorwart2009Aug,Kaer2010Apr,Madsen2011Jun,Groblacher2015Jul,Segal2016May,Potocnik2018Mar}. 
Then, advanced numerical tools are required for the simulation of the dynamics on a classical computer \cite{Tanimura2020Jul,Wang2008Nov,Hartmann2017,Strathearn2018Aug,Werther2020May,Kundu2023Jun}.

Many of the most sophisticated approaches realize the open system evolution by substituting the original environment with few physical or non-physical auxiliary degrees of freedom.
These auxiliary degrees of freedom must be carefully tailored to accurately reproduce the dynamics of the original bath. Prominent methods in this category include the well established HEOM (hierarchical equations of motion) \cite{Tanimura2020Jul}, HOPS (hierarchy of pure states) \cite{suess14,Flannigan2021Aug} and pseudomode approaches \cite{Imamoglu1994Nov,Garraway1997Mar,Mascherpa2020May}, among others \cite{Hughes2009Jul,Prior2010Jul,Lambert2019Aug}.
However, identifying suitable auxiliary environments is generally a complex task that depends nontrivially on the specific characteristics of the bath structure \cite{Tang2015Dec,Hartmann2019Jun,Xu2022Nov}. A different strategy to treat open system dynamics avoids this issue by working directly with the exact influence functional \cite{Feynman1963Oct}. Viewed as a process tensor, it encapsulates all dynamical properties of the reduced dynamics \cite{Pollock2018Jan}. This tensor has a representation as a two-dimensional tensor network \cite{Makarov1994Apr,Strathearn2018Aug,Jorgensen2019Dec,Otterpohl2022Sep}, which can be contracted to matrix product operator (MPO) form to allow for efficient computations \cite{Jorgensen2019Dec,Fux2021May, Fux2022Jan,Fowler-Wright2022Oct,Gribben2022Feb,Otterpohl2022Sep}. MPO methods are also used widely in the context of weakly dissipative open systems with spatial correlations (see, e.g. \cite{Verstraete2004Nov,Zwolak2004Nov}). In contrast, here, the MPO encodes temporal correlations due to time-nonlocal dynamics induced by a structured bath. 

In this paper we establish an alternative representation of the process tensor in terms of an infinite tensor network. 
This key result allows us to use infinite time evolving block decimation (iTEBD) \cite{Vidal2007Feb} for network contraction, leading to a fast algorithm with a previously unachieved numerical scaling (linear in the bath memory time). The resulting MPO representation of the process tensor has the same structure as for methods using auxiliary degrees of freedom, bridging a gap between the two different approaches. Crucially, this delivers a single time-local propagator, encoding the full dynamics of the open system. Thus, we can reach arbitrary evolution times straightforwardly, and even utilize spectral theory in order to determine stationary states and characterize asymptotic behavior.
In contrast to established methods such as HEOM, the auxiliary degrees of freedom are generated automatically in an optimized and systematic way by the network contraction algorithm.

\paragraph{Open system evolution}
As a model for open system dynamics we consider the standard Hamiltonian
\begin{equation}\label{eq:hamilt}
    H(t)=H_\mathrm{sys}(t)\otimes \id_\mathrm{env}+S\otimes B(t),
\end{equation}
where $H_\mathrm{sys}$ and $S$ are hermitian operators in the Hilbert space of the system and $B(t)$ is an operator that describes the collective degrees of freedom of a Gaussian environment
consisting of a continuum of bosonic modes \cite{deVega2017Jan}. 
This operator is characterized by the so-called bath correlation function
$\alpha(t,s)=\tr \,\rho_\mathrm{env}(0)B(t)B(s)$,
where $\rho_\mathrm{env}(0)$ is a Gaussian environment initial state \footnote{We assume without loss of generality that $\tr\rho_\mathrm{env}(0)B(t)=0$.}.
The bath is said to be stationary if the bath correlation function depends only on the time difference $\alpha(t,s)\equiv\alpha(t-s)$. While notable exceptions exist \cite{Link2022Feb}, this is the standard scenario in open system dynamics. 
In order to arrive at a description of the reduced dynamics in terms of the process tensor, one can employ a Trotter splitting of the full unitary time evolution operator in small time steps $\Delta$ \cite{Hatano2005Nov}.
Then, a time-discrete path integral for the dynamics can be derived, in which the influence of the bath is fully captured by the so-called influence functional. In a more general modern open system framework, the influence functional gives rise to the process tensor from which all dynamical properties of the system can be extracted \cite{Pollock2018Jan,Jorgensen2019Dec}.

For clarity we focus only on computing the system density matrix after $N$ time steps $t=N\Delta$. We use a Liouville-space (density matrix space) notation where a single index $\mu\equiv (\mu_l,\mu_r)$ labels a (\emphquote{left} and \emphquote{right}) pair of eigenstates $\ket{\mu_l}$, $\ket{\mu_r}$ of the coupling operator $S$, such that density matrices are denoted as vectors $\rho^\mu=\braket{\mu_l|\rho|\mu_r}$. Thus, if the dimension of the system Hilbert space is $d$, $\mu$ runs from $1$ to $d^2$. The time evolution of the system state $\rho(t)$ can then be expressed in terms of a discrete path integral \cite{feynman63,Makri1995Mar,Makri1995Mar2,Jorgensen2019Dec,Cygorek2021Jan}
\begin{equation} 
\begin{split}\label{eq:time_evo}
&\rho^{\nu_N}(N\!\Delta)=\hspace{-2mm}\sum\limits_{\substack{\mu_1...\mu_N \\ \nu_0...\nu_{N-1}}}\hspace{-2mm}\mathcal{F}_N^{\mu_1...\mu_N}\Big(\prod_{k=1}^N\mathcal{U}_{k}^{\nu_{k-1}\mu_k\nu_{k}}\Big)\rho^{\nu_0}(0).
\end{split}
\end{equation}
We can write this equation pictorially using tensor network notation 
\begin{equation}\label{eq:time_evo_pic}
    \begin{tikzpicture}[inner sep=1mm, x=.7cm,y=.7cm]
    \foreach \j in {1,2,3} {
        \node[tensor2] (\j) at (\j, 0) {$\mathcal{U}_\j\!$};
        \draw[-] (\j) -- (\j, 1);   
    };
    \node[tensor2] (N) at (5.5, 0) {\hspace{-0.25mm}$\mathcal{U}_{\!N}$\!};
    \draw[-] (N) -- (5.5, 1);   
    \node[] (dots) at (4.32,-0.015) {$\cdots$};
    \foreach \j in {2,3} {
        \pgfmathsetmacro{\iminusone}{int(\j-1)}
        \draw[-] (\j) -- (\iminusone);   
    };
    \draw[-] (N) -- (6.25,0); 
    \draw[-] (4.8,0) -- (N); 
    \draw[-] (3) -- (3.7,0); 
    \node[vector] (is) at (-0.2,0) {$\rho(0)$};
    \draw[-] (is) -- (1); 
    \node[vector] (fs) at (-3,0) {$\rho(N\!\Delta)$};
    \draw[-] (fs) -- (-1.78,0); 
    \node at (-1.25,-0.05) {$=$};
    \node[tensor] (t0) at (3.26,1) {$\hspace{1.5cm}\mathcal{F}_N\hspace{1.5cm}$};

\end{tikzpicture}.
\end{equation}
The tensors $\mathcal{U}_k^{\lambda\mu\nu}$ can be seen as unitary channels describing the evolution due to $H_\mathrm{sys}$ for the time step $k$ \cite{supp_mat} and $\mathcal{F}_N^{\mu_1...\mu_N}$ is the time-discrete influence functional, a rank-$N$ tensor accounting for the time-nonlocal effect of the bath. 

Even though for Gaussian baths the influence functional is known analytically, the time evolution according to Eq.~\eqref{eq:time_evo} involves a sum over all elements of $\mathcal{F}_N$ which are exponentially many ($d^{2N}$). Therefore such a direct computation cannot be used in practice.

\begin{figure}[t]
    \includegraphics[]{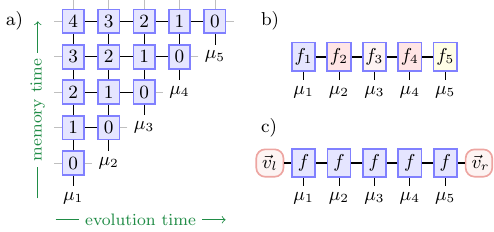}
    \caption{Different tensor network representation of the time discrete influence functional $\mathcal{F}_5^{\mu_1...\mu_5}$ for $N=5$ time steps. a) Exact representation as a two-dimensional tensor network \cite{Jorgensen2019Dec} (greyed out open tensor legs must be summed over). b) Matrix product operator representation as obtained from contraction of the network a) (PT-TEMPO) \cite{Fux2021May}. c) A semi-group representation with identical tensors $f$, as in Eq.~\eqref{eq:F_aux}. }
    \label{fig:my_label}
\end{figure}

\paragraph{Tensor Network Representation of the Influence Functional}
It has been shown in Refs.~\cite{Makarov1994Apr,Strathearn2018Aug,Jorgensen2019Dec} that the time-discrete influence functional can be represented as a two-dimensional tensor network. In detail one can define a set of elementary tensors
\begin{equation}\label{eq:btensor}
    b^{\mu\nu}_{ij}(k)=\begin{cases}
        \delta_{ij} \delta_{\mu\nu} I_k(\mu,j), &k>0\\
        \delta_{ij}  \delta_{\mu\nu}\delta_{j\mu} I_0(\mu,j), &k=0
    \end{cases}
    =  \hspace{-.3cm}
\vcenter{\begin{tikzpicture}[inner sep=1mm, x=.7cm,y=.7cm]
\node[tensor] (b_tens) at (0,0) {$k$};
\node (b) at (0, 1) {$\nu$};
\draw[-] (b_tens) -- (b);    
\node (a) at (0, -1) {$\mu$};
\draw[-] (b_tens) -- (a);    
\node (j) at (0-1, 0) {$j$};
\draw[-] (b_tens) -- (j);    
\node (i) at (0+1, 0) {$i$};
\draw[-] (b_tens) -- (i);     
\end{tikzpicture}}\hspace{-6.7cm}
\end{equation}
such that the influence functional can be expressed as in Fig.~\ref{fig:my_label}a. Here the weights
\begin{equation}
    I_{k}(\mu,\nu)= \exp\left((S_{\mu_l}-S_{\mu_r})(\eta_k S_{\nu_l}-\eta_k^*S_{\nu_r})\right)
\end{equation}
are used, where $\eta_k$ is determined by the bath correlation function at time step $k$ and $S_n$ denotes the $n$'th eigenvalue of the coupling operator \cite{Strathearn2018Aug,Jorgensen2019Dec,supp_mat}. 
In the PT-TEMPO scheme (Process Tensor Time Evolving Matrix Product Operators) \cite{Strathearn2018Aug,Jorgensen2019Dec,Fux2021May,tempoCollaboration2023Jun}, the network (Fig.~\ref{fig:my_label}a) is contracted to a matrix product operator (Fig.~\ref{fig:my_label}b). This can be done for instance by multiplying adjacent columns followed by a compression based on singular value decompositons, which is required to keep the bond dimension manageable. With a MPO form for the influence functional the open system evolution Eq.~\eqref{eq:time_evo},~\eqref{eq:time_evo_pic} can be performed straightforwardly with iterative tensor contractions.
At first sight, to obtain a process tensor for $N$ time steps in MPO form, $\mathcal{O}(N^2)$ matrix factorizations are required to contract the two dimensional network Fig.~\ref{fig:my_label}a. 
However, usually one assumes a finite memory time of the bath such that all $b(k)$ tensors for $k> N_c$ can be neglected ($\eta_{k>N_c}\approx 0$) \footnote{This is no restriction because the system evolution for $N$ time steps only depends on $\alpha(t)$ for $t<N\Delta $. Thus, if the correlation function does not decay, we can add an artificial smooth cutoff to the bath correlation for times after $t=N\Delta$.}.  
In this case it has been shown that the scaling of the network contraction can be improved to $\mathcal{O}(N_c\log N_c)$ \cite{Cygorek2023Apr}. 

In the following we utilize infinite MPO evolution techniques to generate a new MPO representation for the influence functional, taking the form displayed in Fig.~\ref{fig:my_label}c. As a formula we can express this as
\begin{equation}\label{eq:F_aux}
\mathcal{F}_N^{\mu_1...\mu_N}=\vec{v}_l^T f^{\mu_1}f^{\mu_2}\cdots f^{\mu_N}\vec{v}_r 
\end{equation}
where, for given index $\mu$, $f^\mu$ is a square matrix (dimensions $\chi\times\chi$) and $\Vec{v}_{l/r}$ are vectors realizing finite-time boundary conditions. Unlike for the MPO resulting from PT-TEMPO (Fig.~\ref{fig:my_label}b), the tensors $f$ are all identical and independent of $N$, such that the MPO can be trivially extended to arbitrary evolution times. We will later exploit the crucial advantages of this semi-group structure in example calculations. As an important side remark, note that, when using auxiliary degrees of freedom to effectively describe the open system evolution, the time-discrete influence functional also takes the from of Eq.~\eqref{eq:F_aux}. For instance, using the hierarchical equations of motion (HEOM) approach, the tensor $f$ becomes the propagator of the hierarchy for a time step $\Delta$ and the bond dimension $\chi$ is the number of auxiliary density operators \cite{supp_mat}. However, in order to generate the HEOM propagator one has to manually tailor a suitable auxiliary environment. In contrast, our new scheme automatically generates this form in an optimized way based on MPO compression. 

As a first step we expand the exact network Fig.~\ref{fig:my_label}a by extending the index dimension ($d^2$) of the tensors $b(k)$ by one, introducing a \emphquote{zero} dimension via $I_k(0,i)=I_k(i,0)\equiv 1$, and keeping the definition \eqref{eq:btensor} as is. This additional dimension is used only to realize finite-size boundary conditions (boundary vectors in \eqref{eq:F_aux}) and can be discarded later. 
If one index of an extended $b(k)$ tensor is zero, the tensor reduces to a trivial product of delta functions. As demonstrated in the supplementary material \cite{supp_mat}, this property allows us to obtain the influence functional for $M<N$ time steps from the influence functional for $N$ time steps by inserting zeros at the boundary
\begin{equation}\label{eq:F_reduction1}
\mathcal{F}_M^{\mu_1...\mu_M}=\mathcal{F}_N^{0...0,\mu_1...\mu_M}=\mathcal{F}_N^{\mu_1...\mu_M,0...0}.
\end{equation}
It is even possible to factor the influence functional into two, by piercing the train of indices with at least $N_c$ zeros
\begin{equation}\label{eq:F_reduction2}
\mathcal{F}^{\mu_1...\mu_M,0...0,\nu_1...\nu_K}_N=\mathcal{F}_M^{\mu_1...\mu_M}\mathcal{F}_K^{\nu_1...\nu_K}.
\end{equation}
These relations can be used to obtain $\mathcal{F}_N$ from an influence functional $\mathcal{F}_\infty$ with infinite time steps. In fact, infinite tensor network contraction methods allow us to obtain an MPO expression for such an infinite influence functional in the form
\begin{equation}\label{eq:F_inf}
    \mathcal{F}^{...\mu\nu\delta...}_\infty=\tr\left[ \cdots f^\mu f^\nu f^\delta \cdots\right]
\end{equation}
where $f^\mu$ are $\chi \times \chi$ matrices (bond dimension $\chi$, $\mu=0,1,...,d^2$). As the boundary condition for the infinite network is irrelevant, we have chosen periodic boundary conditions for convenience. Using \eqref{eq:F_reduction1} and \eqref{eq:F_reduction2} the desired influence functional for $N$ steps can then be obtained via
\begin{equation}
    \mathcal{F}_N^{\mu_1...\mu_N}=\tr\left[(f^{0})^{\infty}  f^{\mu_1}f^{\mu_2}\cdots f^{\mu_N} \right].
\end{equation} 
The infinite matrix power can be expressed as $(f^0)^\infty=\vec{v}_r\circ\vec{v}_l$ with $\vec{v}_{l/r}$ the leading left and right eigenvectors of $f^0$ (eigenvalue one). We have indeed recovered a representation of the type \eqref{eq:F_aux}.
Crucially, one only needs to compute and store the single tensor $f$ instead of $\mathcal{O}(N_c)$ such tensors as in the finite contraction schemes \cite{Cygorek2023Apr}.  Moreover, as demonstrated in the supplementary material, the stationary state can also be determined efficiently by computing the leading eigenvector of the full short-time propagator  $\mathcal{Q}_{(\lambda,i)}^{(\nu,j)}= \sum_\mu f^\mu_{ij} \mathcal{U}^{\lambda\mu\nu}$ \cite{supp_mat}.

\begin{figure}[t]
    \centering
    \includegraphics[trim={0 0 0 0.68cm},clip=true]{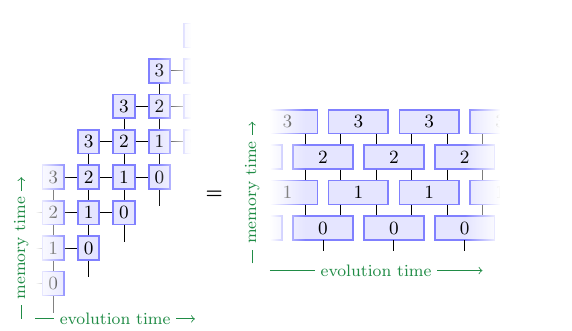}
    \caption{Visual representation of an infinite time-translationally invariant influence functional with $N_c=3$ (left network) which can be seen as nearest-neighbor matrix product operator evolution in memory time (right network).}
    \label{fig:tebd_network}
\end{figure}

\paragraph{Algorithm}
It remains to provide an algorithm for computing the tensor $f$ in Eq.~\eqref{eq:F_inf}. Based on an infinite-$N$ limit of the exact network in Fig.~\ref{fig:my_label}a, we propose a network contraction in an \emphquote{anti-diagonal} direction starting from $k=N_c$, as shown in Fig.~\ref{fig:tebd_network}, when the network displays a structure suitable for time evolving block decimation (TEBD). The \emphquote{gates} $b(k)$ can formally be seen as nearest neighbor coupling alternating between left and right \emphquote{sites}. Thus, it is straightforward to apply infinite TEBD algorithms \cite{Vidal2007Feb,Orus2008Oct,link_github} with MPO evolution from top to bottom (Fig.~\ref{fig:tebd_network} right panel). This requires only $N_c$ matrix factorizations. Since the gates $b(k)$ become weakly entangling for large $k$, the bond dimension increases significantly only for the last few evolution steps, making this an excellent contraction scheme. 
The simple iTEBD algorithm from Ref.~\cite{Orus2008Oct} already performs very well, resulting in similar bond dimension for a given accuracy as the contraction of the finite network, but with a computational speedup in orders of magnitude (for more details on this computational advantage see the supplemental material \cite{supp_mat}). 

\paragraph{Applications} 
For the following examples we consider a (sub-)ohmic bath with exponential cutoff. At zero temperature the bath correlation function reads \cite{Hartmann2017}
\begin{equation}\label{eq:BCF_ohmic}
    \alpha(t)=\alpha \omega_c^{2}  \frac{\Gamma(s + 1)}{2(1 + \ii\omega_c t)^{s+1}}  .
\end{equation}
In this expression, $\alpha$ is a dimensionless coupling strength, $\omega_c$ is the cutoff frequency, and $s\leq 1$ is the exponent of the low frequency behavior $\propto \omega^s$ of the spectral density.
This function decays algebraically for large times, possibly making it challenging for simulations due to a resulting long memory time. 

\begin{figure}
    \centering
    \includegraphics[width=.45\textwidth,trim={0 .2cm 0 0cm},clip=True]{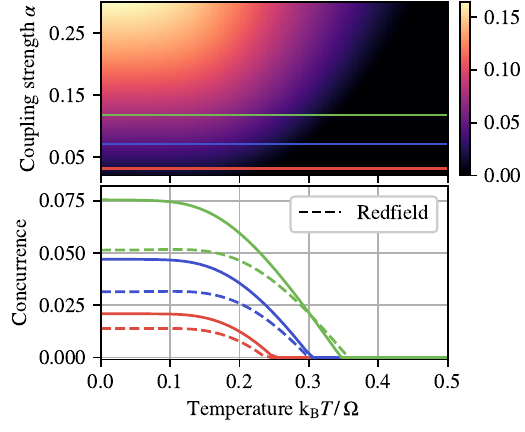}
    \caption{Steady state concurrence in the two spin boson model with an ohmic bath $s=1$ and $\omega_c=5\Omega$ for different coupling strengths and temperatures computed with our new approach (converged results). The lower panel shows cuts for the coupling strengths indicated by the lines in the upper panel. As a comparison, the concurrence predicted by Redfield theory is displayed as dashed lines.}
\label{fig:2sb}
\end{figure}

As a first example we compute the asymptotic entanglement in a two-spin boson model. The model consists of noninteracting spins $A$ and $B$ that are coupled to the same bath via
\begin{equation}\label{eq:2SB_H}
        H(t)=\frac{\Omega}{2} (\sigma_x^A+\sigma_x^B)\otimes \id_\mathrm{env}+\frac{1}{2}\left(\sigma_z^A+ \sigma_z^B\right)\otimes B(t).
\end{equation}
Even if the spins are not directly coupled, at low temperatures, they still become entangled via the interaction with a common bath \cite{Thorwart2009Aug,Hartmann2020Oct}. 
A crucial advantage of our framework is that we can use a spectral decomposition of $\mathcal{Q}$ to determine the steady state without relying on time evolution \cite{Minganti2018Oct,Debecker2023Jul}. This allows us to effortlessly obtain accurate values for the asymptotic concurrence over large parameter regimes, displayed in Fig.~\ref{fig:2sb}. As can be expected, the concurrence increases with the increasing coupling strength and decreases with increasing temperature. For every coupling strength a maximum temperature exists after which the asymptotic state becomes separable. 
Note that even for weak coupling the concurrence is difficult to compute using standard perturbative master equations. As shown in Fig.~\ref{fig:2sb}, the second order Redfield equation \cite{Hartmann2020} predicts systematically wrong values for weak coupling. In fact, second order master equations predict the steady state only to zeroth order accuracy \cite{Fleming2011Mar,Hartmann2020Oct,Crowder2023Oct}, while obtaining higher order equations is tedious \cite{Palacino2021Jul,Crowder2023Oct}.

\begin{figure}
    \centering
    \includegraphics[width=.45\textwidth]{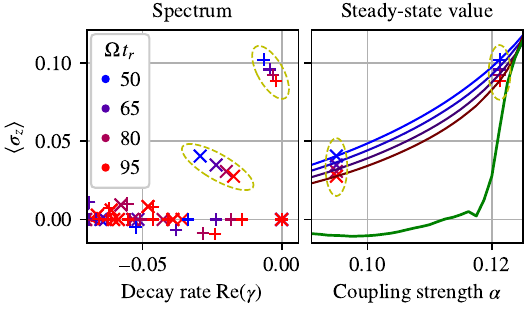}
    \caption{Predictions for the stationary $\sigma_z$ expectation value in the sub-ohmic spin boson model ($s=0.5$, $\omega_c=20\Omega$) using different low frequency regularizations $t_r$. The left panel shows the spectral contributions to $\sigma_z$ from the numerically computed short time propagator below the transition ($\times$ marker) and above the transition ($+$ marker). We can identify the unique steady state ($\gamma=0$, $\braket{\sigma_z}=0$) as well as the next-to-leading contribution which breaks the symmetry (encircled). On the right hand side the predicted steady state values are displayed as a function of the coupling strength. The green line shows the extrapolated values from algebraic fits \cite{supp_mat}. We can identify the phase transition at $\alpha\approx 0.1175$. }
    \label{fig:critical_coupling}
\end{figure}

We further exemplify the power of a spectral analysis by studying the well-known quantum phase transition in the sub-Ohmic spin boson model \cite{Bulla2003Oct,Vojta2005Feb,Anders2007May,Wang2010May,Nalbach2010Feb,Nalbach2013Jan,Ren2022Nov,Xu2022Nov}
\begin{equation}
    H(t)=\Omega \sigma_x\otimes \id_\mathrm{env}+\sigma_z\otimes B(t).
\end{equation}
As the coupling strength $\alpha$ is increased, the system changes from a symmetric phase, where asymptotically $\braket{\sigma_z}=0$, to a symmetry broken phase $\braket{\sigma_z}\neq 0$. In general, such phase transitions are difficult to describe via time evolution because it is hard to separate asymptotic and transient behavior, especially since numerical approaches will typically generate a gapped spectrum \cite{Strathearn2018Aug, Wang2019Mar}. In our framework we can employ a spectral decomposition in order to write the evolution of any observable as
\begin{equation}
    \braket{\sigma_z}(t)=\sum_{k=1}^{\chi d^2}\eul^{\gamma_k t} \braket{\sigma_z}_k,
\end{equation}
where $\gamma_k$ are complex rates extracted from the eigenvalues of the short time propagator $\mathcal{Q}$. For large $t$ we can keep only the two most relevant contributions in the sum, the leading and next-to-leading eigenvector
\begin{equation}
    \braket{\sigma_z}(t)\rightarrow \eul^{\gamma_1 t} \braket{\sigma_z}_1+\eul^{\gamma_2 t} \braket{\sigma_z}_2
\end{equation}
For the spin boson model this requires further justification, because the exact spectrum is not gapped. We provide a full discussion of the subtleties in the supplement \cite{supp_mat}. There always exists a unique steady state contribution with $\gamma_1=0$ which obeys the symmetry of the model $\braket{\sigma_z}_1=0$. Hence, to describe the transition, we must consider the next-to-leading eigenvector. 
In order to ensure convergence of the algorithm, we modify the bath correlation function after a time $t_r$ to decay exponentially (low frequency regularization). 
The original spin boson model is recovered when $t_r\rightarrow\infty$. In this limit we find for all coupling strengths that $\gamma_2\rightarrow 0$ (see Fig.~\ref{fig:critical_coupling}). Thus, the symmetry breaking is characterized by the value of $\braket{\sigma_z}_2$ extrapolated to large $t_r$. Since the phase transition is of second order, we make an extrapolation by fitting algebraic curves to the numerical data. The results are displayed in Fig.~\ref{fig:critical_coupling}. While the curves for a finite $t_r$ (red and blue) do not indicate the transition point, we can clearly identify the critical coupling from the extrapolated values (green curve). 

\paragraph{Conclusions}

Matrix product operators have proven to be highly efficient in representing the temporal correlations (memory) in the quantum evolution of open systems \cite{Strathearn2018Aug,Fux2021May,Cygorek2021Jan,Otterpohl2022Sep}. These correlations are encoded in the influence functional (or the process tensor) which can be seen as a MPO in evolution time \cite{Jorgensen2019Dec}. 
In this work we have demonstrated that, within this framework, a MPO form of the influence functional for arbitrary finite or infinite evolution times can be obtained by contraction of a single infinite tensor network. This strategy has crucial advantages over previous approaches that were based on finite tensor network contractions \cite{Jorgensen2019Dec,Cygorek2023Apr}. The new contraction algorithm achieves an optimal scaling with respect to the number of required matrix operations $\mathcal{O}(N_c)$ and leads to a considerable computational speed-up over all previously known approaches which require at least $\mathcal{O}(N_c\log N_c)$ ($N_c$ is the number of memory time steps) \cite{Jorgensen2019Dec,Cygorek2023Apr}. Even more significantly, we obtain a single time-independent (semi-group) propagator that delivers the full open system evolution. This structural advantage allows us to utilize a spectral decomposition in order to characterize particularly relevant asymptotic dynamics even in difficult settings such as dissipative phase transitions. From a broader perspective, our result can be seen as a way to automatically generate an optimized set of auxiliary degrees of freedom which realize the exact bath response to a controlled level of accuracy.
While the simple iTEBD algorithm that we use here performs very well already, we believe there is substantial potential for further optimization. For instance, using advanced infinite MPS evolution schemes~\cite{Zauner-Stauber2018Jan,Vanderstraeten2019Jan,Vanhecke2021Feb} could lead to a better accuracy at a given bond dimension, which becomes relevant for large system sizes and ultra strong coupling. Moreover, we are hopeful that similar schemes can be developed for more general couplings \cite{Gribben2022Feb} as well as fermionic \cite{Ng2023Mar} and non-Gaussian baths \cite{Cygorek2021Jan}.

\begin{acknowledgments}
\paragraph{Acknowledgements} V.L.~and W.T.S.~gratefully acknowledge discussions with Richard Hartmann concerning the numerical examples.  We also thank Kai Müller, Jonathan Keeling, and anonymous Referees for their valuable comments.
H.-H.T. is supported by the Deutsche Forschungsgemeinschaft (DFG) through project A06 of SFB 1143 (project No.~247310070).
\end{acknowledgments}

\bibliography{bib.bib}

\clearpage
\onecolumngrid

\section*{Supplementary Material}
\renewcommand\thefigure{S\arabic{figure}}    
\setcounter{figure}{0}    
\renewcommand\theequation{S\arabic{equation}}    
\setcounter{equation}{0}

\subsection{A. Open system dynamics with the influence functional}

In order to arrive at a description of the dynamics in terms of the influence functional, we can consider a Trotter splitting of the full unitary time evolution operator \cite{Hatano2005Nov}
\begin{equation}
    U(t+\Delta,t) = U_\mathrm{sys}(t+\Delta,t+\Delta/2)U_\mathrm{int}(t+\Delta,t)U_\mathrm{sys}(t+\Delta/2,t) +\mathcal{O}(\Delta^3),
\end{equation}
where $U_\mathrm{sys}(t,s)$ is the unitary evolution operator generated by $H_\mathrm{sys}$ and $U_\mathrm{int}(t,s)$ is the evolution operator generated by the interaction term in Hamiltonian \eqref{eq:hamilt}.
With the standard steps for the derivation of a path integral one can obtain an expression for the reduced state of the system 
\begin{equation}
    \rho(t)=\tr_\mathrm{env}U(t,0)\rho(0)\otimes\rho_\mathrm{env}(0)U^\dagger(t,0)
\end{equation}
in the form (Eq.~\eqref{eq:time_evo} main text, repeated here for convenience) \cite{feynman63,Makri1995Mar,Makri1995Mar2,Jorgensen2019Dec,Cygorek2021Jan}
\begin{equation} 
\begin{split}\label{eq:time_evo_supp}
&\rho^{\nu_N}(N\!\Delta)=\hspace{-2mm}\sum\limits_{\mu_1...\mu_N=1}^{d^2}\sum\limits_{\nu_0...\nu_{N-1}=1}^{d^2}\mathcal{F}_N^{\mu_1...\mu_N}\Big(\prod_{k=1}^N\mathcal{U}_{k}^{\nu_{k-1}\mu_k\nu_{k}}\Big)\rho^{\nu_0}(0).
\end{split}
\end{equation}
The tensors $\mathcal{U}$ describe purely unitary evolution and are defined as
\begin{equation}
    \mathcal{U}_{k}^{\lambda\mu\nu}={u}^{\nu\mu}_\mathrm{sys}(k\Delta+\Delta,k\Delta+\Delta/2){u}^{\mu\lambda}_\mathrm{sys}(k\Delta+\Delta/2,k\Delta)
\end{equation}
where $u^{\mu\nu}_\mathrm{sys}(t,s)$ is the unitary channel for the evolution due to $H_\mathrm{sys}$ from time $s$ to $t$
\begin{equation}
        u_\mathrm{sys}^{\mu\nu}(t,s)=\braket{\mu_l|U_\mathrm{sys}(t,s)|\nu_l}\braket{\nu_r|U_\mathrm{sys}^\dagger(t,s)|\mu_r}.
\end{equation}
For Gaussian bosonic baths the influence functional takes the exact form \cite{feynman63,Makri1995Mar,Makri1995Mar2,Jorgensen2019Dec}
\begin{equation}\label{eq:f_stat}
\begin{split}
&\mathcal{F}_N^{\mu_1...\mu_N}=\prod_{i=1}^N\prod_{j=1}^i I_{(i-j)}(\mu_i,\mu_j)
\,,\qquad I_{k}(\mu,\nu)= \exp\left(-(S_{\mu_l}-S_{\mu_r})(\eta_k S_{\nu_l}-\eta_k^*S_{\nu_r})\right),
\end{split}
\end{equation}
where the discretized bath correlation function is given as
\begin{equation}
    \eta_{k}=  \begin{cases}
      \int_{k\Delta}^{(k+1)\Delta}\diff t\int_{0}^{\Delta}\diff s\,\alpha(t-s), & k>0 \\
      \int_{0}^{\Delta}\diff t\int_{0}^{t}\diff s\,\alpha(t-s), & k=0
    \end{cases}
\end{equation}
and $S_n$ denotes the $n$'th eigenvalue of the coupling operator $S$.

In the following we assume a time-independent system Hamiltonian such that the unitary evolution is independent of the time step $\mathcal{U}_k^{\lambda\mu\nu}\equiv \mathcal{U}^{\lambda\mu\nu}$.
Further, we assume that we have an MPO representation of the influence functional in the form Eq.~\eqref{eq:F_aux} (main text). A crucial advantage of the resulting semi-group formulation of the dynamics is that we can study low-frequency asymptotic dynamics by looking at the spectral properties of the short-time propagator \cite{Debecker2023Jul}. This propagator can be defined as
\begin{equation}
    \mathcal{Q}_{(\lambda,i)}^{(\nu,j)}=  \sum_{\mu=1}^{d^2}\mathcal{U}^{\lambda\mu\nu}\,f^\mu_{ij}.
\end{equation}
Following Eq.~\eqref{eq:time_evo_supp}, this object can be used to compute the open system evolution for any time $t$
\begin{equation}
    \rho^{\nu}(t)=\sum_{\lambda=1}^{d^2}\sum_{i,j=1}^\chi\,(\mathcal{Q}^{t/\Delta})_{(\lambda,i)}^{(\nu,j)}\,\rho_\mathrm{sys}^\lambda(0)v_l^iv_r^j.
\end{equation}
If we are interested in asymptotic dynamics (large $t$) we should consider the eigendecomposition of $\mathcal{Q}$ in order to evaluate the matrix power $\mathcal{Q}^{t/\Delta}$. Since $\mathcal{Q}$ is non-Hermitian its spectral decomposition takes the form
\begin{equation}
    \mathcal{Q}_{(\lambda,i)}^{(\nu,j)}=\sum_{k=1}^{\chi d^2} q_k (l_k)^{(\nu,j)}(r_k)_{(\lambda,i)},
\end{equation}
with complex eigenvalues $q_k\in \mathbb{C}$ and mutually orthogonal left and right eigenvectors $l_m^\dagger r_n=\delta_{mn}$. This way we can decompose the system evolution as
\begin{equation}\label{eq:spectral_decomp}
        \rho_\mathrm{sys}^{\nu}(t)=\sum_{k=1}^{\chi d^2} q_k^{t/\Delta}\rho_k^\nu=\sum_{k=1}^{\chi d^2} \eul^{\gamma_k t}\rho_k^\nu,\,\qquad \rho_k^\nu=\sum_{\lambda=1}^{d^2}\sum_{i,j=1}^\chi\,(l_k)^{(\nu,j)}(r_k)_{(\lambda,i)} \,\rho_\mathrm{sys}^\lambda(0)v_l^iv_r^j.
\end{equation}
We have defined the rates corresponding to each eigenvalue as $\gamma_k = \log(q_k)/\Delta$. Since the density operator is finite for any time we must have $\mathrm{Re}(\gamma_k)\leq 0$. The asymptotic properties of the model are determined by the behavior of this spectrum close to zero. The unique steady state of a system is the state $\rho_1$ where $\gamma_1=0$. As is known from the spectral theory of Lindblad equations \cite{Minganti2018Oct}, one has $\tr\rho_1=1$ whereas $\tr\rho_k=0$ for all other terms.

\subsection{B. MPO representation of the influence functional from HEOM}
We show how Eq.~\eqref{eq:F_aux} can be obtained from open system dynamics with auxiliary environments, considering HEOM as an example \cite{Tanimura2020Jul}. In the standard HEOM scheme, the bath correlation function is represented approximately as a sum over few exponentials, \cite{Xu2022Nov}
\begin{equation}
 \alpha(t)\approx\sum_{j=1}^MG_j\eul^{-\mathrm{Re}W_j|t|-\ii \mathrm{Im}W_j t},
\end{equation}
with complex parameters $G_j,W_j\in \mathbb{C}$. One then defines an infinite set of auxiliary density operators labeled by a pair of multiindices $\boldsymbol{n},\boldsymbol{m}\in \mathbb{N}_0^M$. These auxiliary states satisfy the hierarchical equation of motion
\begin{equation}\label{eq:HEOM}
    \begin{split}
        \partial_t\rho^{(\bvec{n},\bvec{m})}=& -i[H_\mathrm{sys}, \rho^{(\bvec{n},\bvec{m})}] - \left(\bvec{W}\cdot\bvec{n}+\bvec{W}^*\cdot\bvec{m}\right)\rho^{(\bvec{n},\bvec{m})}+\sum_{j}\left(G_jn_jS\rho^{(\bvec{n}-\bvec{e}_j,\bvec{m})} + G_j^*m_j\rho^{(\bvec{n},\bvec{m}-\bvec{e}_j)}S\right)\\
        &+\sum_{j}\left[\rho^{(\bvec{n}+\bvec{e}_j,\bvec{m})}, S\right] + \left[S, \rho^{(\bvec{n},\bvec{m}+\bvec{e}_j)}\right]\!.
    \end{split}
\end{equation}
We denote $\boldsymbol{e}_i$ the unit vector in direction $i$.
Choosing as an initial state $\rho^{(\bvec{0},\bvec{0})}(0)=\rho(0)$ and all other states zero, one can obtain the reduced system evolution from
$\rho(t)=\rho^{(\bvec{0},\bvec{0})}(t)$. In practice, the hierarchy is cut at sufficiently high index values such that a finite system can be evolved numerically. We can reformulate the hierarchy by embedding it as an operator in an extended Hilbertspace $\rho^{(\bvec{n},\bvec{m})}\equiv\braket{\bvec{n}|R|\bvec{m}}$ \cite{Flannigan2021Aug,Gao2021Sep}. Defining raising and lowering operators $A_i^\pm\ket{\boldsymbol{n}}=\ket{\boldsymbol{n}\pm\boldsymbol{e}_i}$, and counting operators $N_i\ket{\boldsymbol{n}}=n_i\ket{\boldsymbol{n}}$ the hierarchy is mapped onto
\begin{equation}
    \begin{split}
        \partial_tR&= -i[H_\mathrm{sys}, R] - \left(\bvec{W}\cdot\bvec{N}R+R\bvec{W}^*\cdot\bvec{N}\right)+\sum_{j}\left(G_jA^+_jN_jSR+ G_j^*RSN_j A^-_j\right)+\sum_{j}\big(\left[A^-_jR, S\right] + \left[S, RA^+_j\right]\big)
        \\&\equiv \mathcal{L}_\mathrm{sys} R + \mathcal{L}_\mathrm{int} R\,.
    \end{split}
\end{equation}
Since the evolution is linear we can employ a Trotter splitting with time step $\Delta$ between $\mathcal{L}_\mathrm{sys}=-\ii [H_\mathrm{sys},\cdot]$ and $\mathcal{L}_\mathrm{int}$. $\mathcal{L}_\mathrm{int}$ commutes with the coupling operator $S$, so we consider the action of this generator on a product of $S$-eigenstates $\ketbra{\mu_l}{\mu_r}$ labeled with the index $\mu=(\mu_l,\mu_r)$
\begin{equation}
    \mathcal{L}_\mathrm{int}\left(\ketbra{\mu_l}{\mu_r}\otimes x\right)=\ketbra{\mu_l}{\mu_r}\otimes\mathcal{L}_\mathrm{int}^\mu x.
\end{equation}
The generator of the hierarchy dynamics conditioned on the system basis state is given as
\begin{equation}
    \mathcal{L}_\mathrm{int}^\mu x= - \left(\bvec{W}\cdot\bvec{N}x+x\bvec{W}^*\cdot\bvec{N}\right)+\sum_{j}\left(G_jA^+_jN_jS_{\mu_l} x+ G_j^*xS_{\mu_r} N_j A^-_j\right)+\sum_{j}\big(A^-_jx (S_{\mu_r}-S_{\mu_l}) + (S_{\mu_l}-S_{\mu_r})xA^+_j\big) .
\end{equation}
The tensor $f^\mu$ is just the propagator generated by $\mathcal{L}_\mathrm{int}^\mu$ for a time step $\Delta$
\begin{equation}
    f^\mu=\eul^{\Delta \mathcal{L}_\mathrm{int}^\mu}.
\end{equation}
The bond dimension of $f$ is exactly the number of auxiliary density operators that are taken into account in the hierarchy. We obtain an influence functional of the form \eqref{eq:F_aux} with the boundary vectors $v_l=v_r=\ketbra{\boldsymbol{0}}{\boldsymbol{0}}$ (the system state is given by $\rho=\braket{\boldsymbol{0}|R|\boldsymbol{0}}$).

\subsection{C. Reduction of the influence functional}

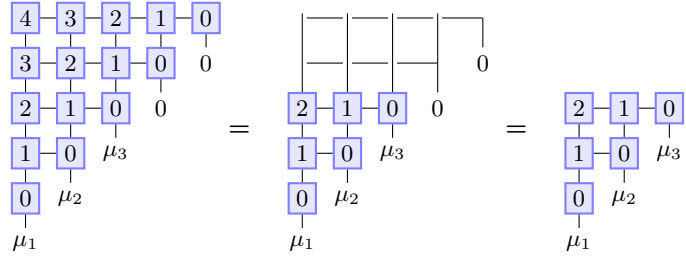
\begin{figure}[h]
    \centering
\begin{tikzpicture}[inner sep=1mm, x=.6cm,y=.6cm]
    \foreach \j in {1,...,3} {
    \foreach \i in {1,...,\j} {
        \pgfmathsetmacro{\jminusi}{int(\j-\i)}        
        \node[tensor] (\i \j) at (\i, \j) {\jminusi};
    };
            \node (\j index) at (\j, \j-1) {$\mu_\j$};
            \draw[-] (\j \j) -- (\j index);
    };
    \foreach \j in {4,...,5} {
    \foreach \i in {1,...,\j} {
        \pgfmathsetmacro{\jminusi}{int(\j-\i)}        
        \node[tensor] (\i \j) at (\i, \j) {\jminusi};
    };
            \node (\j index) at (\j, \j-1) {$0$};
            \draw[-] (\j \j) -- (\j index);
    };

    \foreach \j in {1,...,4} {
    \foreach \i in {1,...,\j} {
        \pgfmathsetmacro{\jplusone}{int(\j+1)}
        \draw[-] (\i \j) -- (\i \jplusone);       
    };
    };
    \foreach \i in {2,...,5} {
    \foreach \j in {\i,...,5} {
        \pgfmathsetmacro{\iminusone}{int(\i-1)}
        \draw[-] (\i \j) -- (\iminusone \j);       
    };
    };
    \node[text width=3cm] at (8,2.5) {$\boldsymbol{=}$};

\end{tikzpicture}\hspace{-2.5cm}
\begin{tikzpicture}[inner sep=1mm, x=.6cm,y=.6cm]
    \foreach \j in {1,...,3} {
    \foreach \i in {1,...,\j} {
        \pgfmathsetmacro{\jminusi}{int(\j-\i)}        
        \node[tensor] (\i \j) at (\i, \j) {\jminusi};
    };
            \node (\j index) at (\j, \j-1) {$\mu_\j$};
            \draw[-] (\j \j) -- (\j index);
    };

    \foreach \j in {1,...,2} {
    \foreach \i in {1,...,\j} {
        \pgfmathsetmacro{\jplusone}{int(\j+1)}
        \draw[-] (\i \j) -- (\i \jplusone);       
    };
    };
    \foreach \i in {2,...,3} {
    \foreach \j in {\i,...,3} {
        \pgfmathsetmacro{\iminusone}{int(\i-1)}
        \draw[-] (\i \j) -- (\iminusone \j);       
    };
    };
    \foreach \i in {2,...,4} {
        \pgfmathsetmacro{\iminusone}{int(\i-1)}
        \draw[-] (\i-.1, 5) -- (\iminusone+.1, 5);       
    };
    \draw[-] (4+.1, 5) -- (5, 5);      
    \node (zero) at (5, 5-1) {$0$};
    \draw[-] (zero) -- (5, 5);    
    \draw[-] (3+.1, 4) -- (4, 4);      
    \node (zero2) at (4, 4-1) {$0$};
    \draw[-] (zero2) -- (4, 4);
    \draw[-] (4,4) -- (4, 5+.1);

    \foreach \j in {1,...,3} {
        \pgfmathsetmacro{\jplusone}{int(\j+1)}
        \draw[-] ( \j 3) -- (\j,5+.1);       
    };
    \foreach \i in {2,...,3} {
        \pgfmathsetmacro{\iminusone}{int(\i-1)}
        \draw[-] (\i-.1, 4) -- (\iminusone+.1, 4);       
    };
    \node[text width=3cm] at (8,2.5) {$\boldsymbol{=}$};

\end{tikzpicture}\hspace{-2.5cm}
\begin{tikzpicture}[inner sep=1mm, x=.6cm,y=.6cm]
    \foreach \j in {1,...,3} {
    \foreach \i in {1,...,\j} {
        \pgfmathsetmacro{\jminusi}{int(\j-\i)}        
        \node[tensor] (\i \j) at (\i, \j) {\jminusi};
    };
            \node (\j index) at (\j, \j-1) {$\mu_\j$};
            \draw[-] (\j \j) -- (\j index);
    };

    \foreach \j in {1,...,2} {
    \foreach \i in {1,...,\j} {
        \pgfmathsetmacro{\jplusone}{int(\j+1)}
        \draw[-] (\i \j) -- (\i \jplusone);       
    };
    };
    \foreach \i in {2,...,3} {
    \foreach \j in {\i,...,3} {
        \pgfmathsetmacro{\iminusone}{int(\i-1)}
        \draw[-] (\i \j) -- (\iminusone \j);       
    };
    };

\end{tikzpicture} \caption{Reduction of the influence functional for 5 time steps to the influence functional for 3 time steps. Tensor legs that are not drawn have to be summed over.}
    \label{fig:reduction1}
\end{figure}

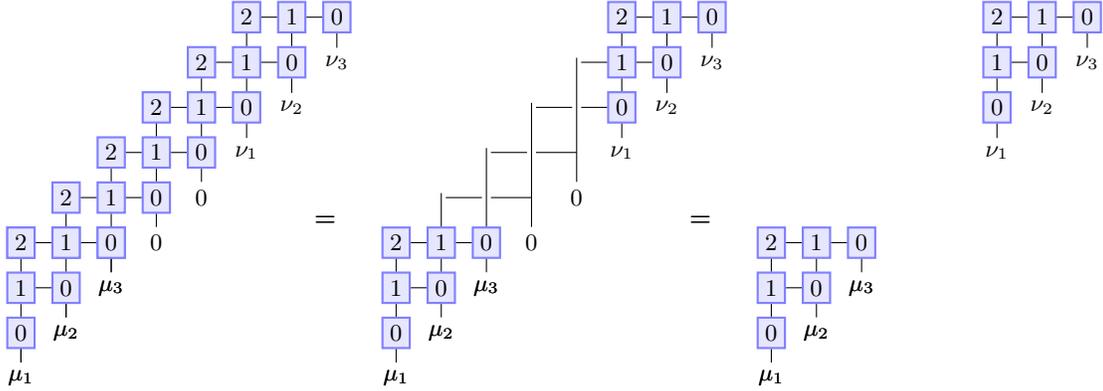
\begin{figure}[h]
    \centering
\begin{tikzpicture}[inner sep=1mm, x=.6cm,y=.6cm]
    \foreach \j in {1,...,3} {
            \pgfmathsetmacro{\jz}{int(\j+3)}

    \foreach \i in {\j,...,3} {
        \pgfmathsetmacro{\jminusi}{int(-\j+\i)}        
        \node[tensor] (\j \i) at (\j, \i) {\jminusi};
    };
            \node (\j index) at (\j, \j-1) {$\mu_\j$};
            \draw[-] (\j \j) -- (\j index);
    };
    \foreach \j in {1,...,2} {
    \foreach \i in {1,...,\j} {
        \pgfmathsetmacro{\jplusone}{int(\j+1)}
        \draw[-] (\i \j) -- (\i \jplusone);       
    };
    };
    \foreach \i in {2,...,3} {
    \foreach \j in {\i,...,3} {
        \pgfmathsetmacro{\iminusone}{int(\i-1)}
        \draw[-] (\i \j) -- (\iminusone \j);       
    };
    };
    
    \foreach \j in {1,...,3} {
            \pgfmathsetmacro{\jz}{int(\j+2)}

    \foreach \i in {3,...,\jz} {
        \pgfmathsetmacro{\jminusi}{int(-\j+\i)}        
        \node[tensor] (\j \i) at (\j, \i) {\jminusi};
    };
            \node (\j index) at (\j, \j-1) {$\mu_\j$};
            \draw[-] (\j \j) -- (\j index);
    };
    \foreach \j in {4,...,5} {
            \pgfmathsetmacro{\jz}{int(\j+2)}

    \foreach \i in {\j,...,\jz} {
        \pgfmathsetmacro{\jminusi}{int(-\j+\i)}        
        \node[tensor] (\j \i) at (\j, \i) {\jminusi};
    };
            \node (\j index) at (\j, \j-1) {$0$};
            \draw[-] (\j \j) -- (\j index);
    };
        \foreach \j in {6,...,8} {
            \pgfmathsetmacro{\jz}{int(\j+3)}
        \pgfmathsetmacro{\jminusseven}{int(\j-5)}        

    \foreach \i in {6,...,\j} {
        \pgfmathsetmacro{\jminusi}{int(\j-\i)}   

        \node[tensor] (\i \j) at (\i, \j) {\jminusi};
    };
            \node (\j index) at (\j, \j-1) {$\nu_\jminusseven$};
            \draw[-] (\j \j) -- (\j index);
    };
    \foreach \j in {6,...,7} {
    \foreach \i in {6,...,\j} {
        \pgfmathsetmacro{\jplusone}{int(\j+1)}
        \draw[-] (\i \j) -- (\i \jplusone);       
    };
    };
    \foreach \i in {7,...,8} {
    \foreach \j in {\i,...,8} {
        \pgfmathsetmacro{\iminusone}{int(\i-1)}
        \draw[-] (\i \j) -- (\iminusone \j);       
    };
    };
        \foreach \i in {3,...,5} {
            \pgfmathsetmacro{\iplusone}{int(\i+1)}
    \foreach \j in {\i,...,\iplusone} {
        \pgfmathsetmacro{\jplusone}{int(\j+1)}
        \draw[-] (\i \j) -- (\i \jplusone);       
    };
    };
    \foreach \i in {4,...,6} {
            \pgfmathsetmacro{\iplusone}{int(\i-1)}
    \foreach \j in {\iplusone,...,\i} {
        \pgfmathsetmacro{\jplusone}{int(\j-1)}
        \draw[-] (\j \i) -- (\jplusone \i);       
    };
    };
    \foreach \i in {2,...,2} {
    \foreach \j in {3,...,3} {
        \pgfmathsetmacro{\jplusone}{int(\j+1)}
        \draw[-] (\i \j) -- (\i \jplusone);       
    };
    };
    \foreach \i in {5,...,5} {
    \foreach \j in {7,...,7} {
        \pgfmathsetmacro{\iplusone}{int(\i+1)}
        \draw[-] (\i \j) -- (\iplusone \j);       
    };
    };    
    \node[text width=3cm] at (10,3.5) {$\boldsymbol{=}$};
    \end{tikzpicture}
\hspace{-2.5cm}
\begin{tikzpicture}[inner sep=1mm, x=.6cm,y=.6cm]
    \foreach \j in {1,...,3} {
            \pgfmathsetmacro{\jz}{int(\j+3)}

    \foreach \i in {\j,...,3} {
        \pgfmathsetmacro{\jminusi}{int(-\j+\i)}        
        \node[tensor] (\j \i) at (\j, \i) {\jminusi};
    };
            \node (\j index) at (\j, \j-1) {$\mu_\j$};
            \draw[-] (\j \j) -- (\j index);
    };
    \foreach \j in {1,...,2} {
    \foreach \i in {1,...,\j} {
        \pgfmathsetmacro{\jplusone}{int(\j+1)}
        \draw[-] (\i \j) -- (\i \jplusone);       
    };
    };
    \foreach \i in {2,...,3} {
    \foreach \j in {\i,...,3} {
        \pgfmathsetmacro{\iminusone}{int(\i-1)}
        \draw[-] (\i \j) -- (\iminusone \j);       
    };
    };
    
    \foreach \j in {1,...,3} {
            \pgfmathsetmacro{\jz}{int(\j+2)}

    \foreach \i in {3,...,\jz} {
        \pgfmathsetmacro{\jminusi}{int(-\j+\i)}        
    };
            \node (\j index) at (\j, \j-1) {$\mu_\j$};
    };
    \foreach \j in {4,...,5} {
            \pgfmathsetmacro{\jz}{int(\j+2)}
    \foreach \i in {\j,...,\jz} {
        \pgfmathsetmacro{\jminusi}{int(-\j+\i)}        
    };
            \node (\j index) at (\j, \j-1) {$0$};
    };
        \foreach \j in {6,...,8} {
            \pgfmathsetmacro{\jz}{int(\j+3)}
        \pgfmathsetmacro{\jminusseven}{int(\j-5)}        

    \foreach \i in {6,...,\j} {
        \pgfmathsetmacro{\jminusi}{int(\j-\i)}   

        \node[tensor] (\i \j) at (\i, \j) {\jminusi};
    };
            \node (\j index) at (\j, \j-1) {$\nu_\jminusseven$};
            \draw[-] (\j \j) -- (\j index);
    };
    \foreach \j in {6,...,7} {
    \foreach \i in {6,...,\j} {
        \pgfmathsetmacro{\jplusone}{int(\j+1)}
        \draw[-] (\i \j) -- (\i \jplusone);       
    };
    };
    \foreach \i in {7,...,8} {
    \foreach \j in {\i,...,8} {
        \pgfmathsetmacro{\iminusone}{int(\i-1)}
        \draw[-] (\i \j) -- (\iminusone \j);       
    };
    };
        \foreach \i in {4,...,5} {
        \draw[-] (\i index) -- (\i, \i+2.1);       
    };
        \pgfmathsetmacro{\i}{int(2)}
        \pgfmathsetmacro{\j}{int(3)}
        \draw[-] (\i \j) -- (\i, \j+1.1);       
        \pgfmathsetmacro{\i}{int(3)}
        \pgfmathsetmacro{\j}{int(3)}
        \draw[-] (\i \j) -- (3, 4+1.1);       

    \draw[-] (5-.1, 6) -- (4+.1, 6);  
    \draw[-] (4-.1, 5) -- (3+.1, 5);  
    
    \draw[-] (5, 5) -- (4+.1, 5);  
    \draw[-] (4, 4) -- (3+.1, 4);  

    \draw[-] (3-.1, 4) -- (2+.1, 4);  
    
    \pgfmathsetmacro{\i}{int(6)}
    \pgfmathsetmacro{\j}{int(6)}
    \draw[-] (\i \j) -- (5+.1, 6);  
    \pgfmathsetmacro{\i}{int(6)}
    \pgfmathsetmacro{\j}{int(7)}
    \draw[-] (\i \j) -- (5+.1, 7);  

    \node[text width=3cm] at (10,3.5) {$\boldsymbol{=}$};
    \end{tikzpicture}
\hspace{-2.5cm}
\begin{tikzpicture}[inner sep=1mm, x=.6cm,y=.6cm]
    \foreach \j in {1,...,3} {
            \pgfmathsetmacro{\jz}{int(\j+3)}

    \foreach \i in {\j,...,3} {
        \pgfmathsetmacro{\jminusi}{int(-\j+\i)}        
        \node[tensor] (\j \i) at (\j, \i) {\jminusi};
    };
            \node (\j index) at (\j, \j-1) {$\mu_\j$};
            \draw[-] (\j \j) -- (\j index);
    };
    \foreach \j in {1,...,2} {
    \foreach \i in {1,...,\j} {
        \pgfmathsetmacro{\jplusone}{int(\j+1)}
        \draw[-] (\i \j) -- (\i \jplusone);       
    };
    };
    \foreach \i in {2,...,3} {
    \foreach \j in {\i,...,3} {
        \pgfmathsetmacro{\iminusone}{int(\i-1)}
        \draw[-] (\i \j) -- (\iminusone \j);       
    };
    };
    
    \foreach \j in {1,...,3} {
            \pgfmathsetmacro{\jz}{int(\j+2)}

    \foreach \i in {3,...,\jz} {
        \pgfmathsetmacro{\jminusi}{int(-\j+\i)}        
    };
            \node (\j index) at (\j, \j-1) {$\mu_\j$};
    };

        \foreach \j in {6,...,8} {
            \pgfmathsetmacro{\jz}{int(\j+3)}
        \pgfmathsetmacro{\jminusseven}{int(\j-5)}        

    \foreach \i in {6,...,\j} {
        \pgfmathsetmacro{\jminusi}{int(\j-\i)}   

        \node[tensor] (\i \j) at (\i, \j) {\jminusi};
    };
            \node (\j index) at (\j, \j-1) {$\nu_\jminusseven$};
            \draw[-] (\j \j) -- (\j index);
    };
    \foreach \j in {6,...,7} {
    \foreach \i in {6,...,\j} {
        \pgfmathsetmacro{\jplusone}{int(\j+1)}
        \draw[-] (\i \j) -- (\i \jplusone);       
    };
    };
    \foreach \i in {7,...,8} {
    \foreach \j in {\i,...,8} {
        \pgfmathsetmacro{\iminusone}{int(\i-1)}
        \draw[-] (\i \j) -- (\iminusone \j);       
    };
    };

\end{tikzpicture}     \caption{Factorization of the influence functional with $N_c=2$ into two independent influence functionals. Tensor legs that are not drawn have to be summed over.}
\label{fig:reduction2}
\end{figure}

We provide a proof for the statements \eqref{eq:F_reduction1} and \eqref{eq:F_reduction2}. These allow us to recover the influence functional for finite times from the infinite version. From our definition, the extended tensors $b(k)$ have the property that, if one index is zero, they reduce to a product of delta functions. In particular
\begin{equation}
    b^{0\nu}_{ij}(k)=\begin{cases}
        \delta_{ij} \delta_{0\nu}, &k>0\\
        \delta_{ij}  \delta_{0\nu}\delta_{j0}, &k=0
    \end{cases}\,,\qquad     b^{\mu\nu}_{0j}(k)=\begin{cases}
        \delta_{0j} \delta_{\mu\nu}, &k>0\\
        \delta_{0j}  \delta_{\mu\nu}\delta_{j\mu}, &k=0
    \end{cases}\,.
\end{equation}
We write this as
\begin{equation}
    b^{\mu\nu}_{0j}(k>0)    = \hspace{-.3cm}
\vcenter{\begin{tikzpicture}[inner sep=1mm, x=.7cm,y=.7cm]
\node (b) at (0, 1) {$\nu$};
\draw[-] (0, 0) -- (b);    
\node (a) at (0, -1) {$\mu$};
\draw[-] (0, 0) -- (a);    
\node (j) at (0-1, 0) {$j$};
\draw[-] (0-.1, 0) -- (j);    
\node (i) at (0+1, 0) {$0$};
\draw[-] (.1, 0) -- (i);     
\end{tikzpicture}}\hspace{-15.5cm} \qquad
    b^{0\nu}_{ij}(k>0)    = \hspace{-.3cm}
\vcenter{\begin{tikzpicture}[inner sep=1mm, x=.7cm,y=.7cm]
\node (b) at (0, 1) {$\nu$};
\draw[-] (0, 0) -- (b);    
\node (a) at (0, -1) {$0$};
\draw[-] (0, 0) -- (a);    
\node (j) at (0-1, 0) {$j$};
\draw[-] (0-.1, 0) -- (j);    
\node (i) at (0+1, 0) {$i$};
\draw[-] (.1, 0) -- (i);     
\end{tikzpicture}}\hspace{-15.5cm} \qquad
    b^{0\nu}_{ij}(0)    = \hspace{-.3cm}
\vcenter{\begin{tikzpicture}[inner sep=1mm, x=.7cm,y=.7cm]
\node (b) at (0, 1) {$\nu$};
\draw[-] (0, 0) -- (b);    
\node (a) at (0, -1) {$0$};
\draw[-] (0, 0) -- (a);    
\node (j) at (0-1, 0) {$j$};
\draw[-] (0, 0) -- (j);    
\node (i) at (0+1, 0) {$i$};
\draw[-] (0, 0) -- (i);     
\end{tikzpicture}}\hspace{-16.5cm}
\end{equation}
The proof of statements \eqref{eq:F_reduction1} and \eqref{eq:F_reduction2} is given pictorially in Figs.~\ref{fig:reduction1} and \ref{fig:reduction2}, respectively.

\subsection{D. Computational advantage}

\begin{figure}
    \centering
    \includegraphics[width=.45\textwidth]{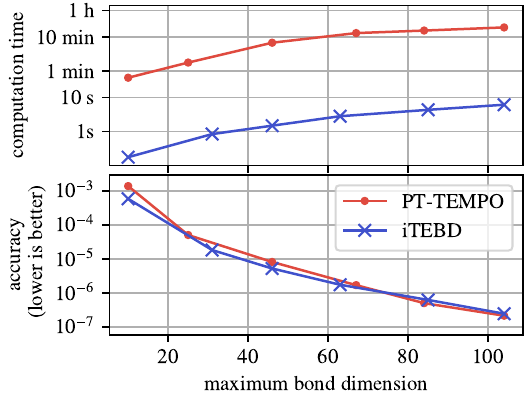}\hspace{.3cm}
    \caption{Contraction of the influence functional of the two-spin boson model in a sub-ohmic bath ($s=0.3$, $\alpha=0.2$). The computation time and the final accuracy of the compressed influence functional is displayed as a function of the bond dimension.
    We chose $N=300$ time steps and $\Delta\omega_c=0.2$. We compare PT-TEMPO from Ref.~\cite{Jorgensen2019Dec} (implementation from~\cite{tempoCollaboration2023Jun}) with our new method using iTEBD \cite{link_github}. Relative error SVD compression was performed using the MKL implementation of LAPACK on consumer hardware. }
    \label{fig:benchmark}
\end{figure}

We provide a benchmark of the iTEBD algorithm for contracting the influence functional. We consider the two-spin boson model \eqref{eq:2SB_H} with sub-ohmic bath \eqref{eq:BCF_ohmic}. In the example Fig.~\ref{fig:benchmark} we show the computation time and the accuracy of the final compressed influence functional for different bond dimensions $\chi$. As a comparison we computed the same problem with the finite contraction scheme from Ref.~\cite{Jorgensen2019Dec} (PT-TEMPO) for $N=N_c=300$. In order to assess the accuracy we consider the average absolute distance of the compressed influence functional to the exact value (from Eq.~\eqref{eq:f_stat}) for a set of $1000$ random \emphquote{paths} (collection of indices $\mu_1,...,\mu_N$).  Our new approach leads to a similar accuracy at a given bond dimension in comparison to PT-TEMPO. However, the iTEBD algorithm requires fewer matrix operations and, moreover, with lower bond dimensions, resulting in a large speedup in the computation time.

\begin{figure}
    \centering
    \includegraphics[width=.45\textwidth]{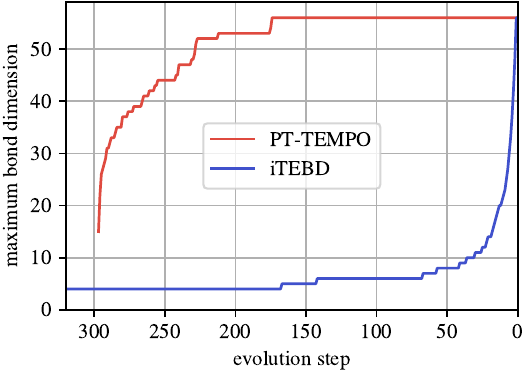}
    \caption{Maximum bond dimension during the network contraction for PT-TEMPO (red) and for the new method based on iTEBD (blue). The system and the parameters are chosen as in Fig.~\ref{fig:benchmark}. Both algorithms generate a MPO representation of the influence functional with maximum bond dimension 56 (final evolution step \emphquote{0}). For the PT-TEMPO simulation we used the code from Ref.~\cite{tempoCollaboration2023Jun} with a relative error SVD truncation threshold of $10^{-7}$. In order to reach the same bond dimension with the iTEBD algorithm we used the threshold $2.8\times 10^{-7}$ (code from Ref.~\cite{link_github}). To ensure a fair comparison, we included an exponential cutoff to the bath correlation function for $t>N\Delta$.  }
    \label{fig:bond_evo}
\end{figure}

We can further analyse the computational advantage of our algorithm over existing proposals. In particular, we can show that the large computational speedup that we observe is not only due to the reduced number of matrix operations, but also due to an optimal network contraction \emphquote{path}. This is an important issue for contracting two-dimensional tensor networks. For instance, while the original PT-TEMPO algorithm (Ref.~\cite{Jorgensen2019Dec}) requires $\mathcal{O}(N^2)$ SVD operations, the divide-and-conquer algorithm presented in Ref.~\cite{Cygorek2023Apr} may not lead to significant speedups even though only $\mathcal{O}(N\log{N})$ SVD operations are required. One reason for this is that the bond dimensions that occur during the divide-and-conquer algorithm are generally larger for the same accuracy threshold, as discussed in Ref.~\cite{Cygorek2023Apr}. 

Our proposed algorithm uses an 'anti-diagonal' contraction from $k=N_c$ to $k=0$. Since the bath correlation function decays for large times, the tensors $b(k)$ are only weakly entangling at large $k$. Hence, the bond dimension is expected to remain small at the beginning of the MPO evolution algorithm. To illustrate this, we display in Fig.~\ref{fig:bond_evo} the bond dimension during the contraction for the two-spin boson model example from the main text (Fig.~\ref{fig:benchmark}). The bond dimension increases very slowly and reaches the final value only at the last evolution step (slow entanglement growth). In contrast, with PT-TEMPO the bond dimension increases fast. The reason for this is that the PT-TEMPO algorithm contracts rows consisting of tensors with all $k$ values in every step. In the example computation, the maximum bond dimension is reached about half-way through the MPO evolution. Thus, SVDs of tensors with this maximum dimension have to be performed for a large number of contraction steps. In contrast, this most costly SVD is performed only once in the iTEBD scheme.

Finally, we note that for the iTEBD algorithm to produce the best results, $N_c$ has to be set large enough for the given SVD truncation threshold. For algebraically decaying bath correlation functions the required $N_c$ value can be controlled by introducing a slow exponential cutoff for large times, which typically does not lower the accuracy of the final MPO representation. In fact, the system dynamics up to time $t_f$ is not influenced by the behavior of the bath correlation function for $t>t_f$. Thus, a cutoff after $t_f$ has no effect on the dynamics. Overall, this issue is not severe because the bond dimensions at the beginning of the contraction are very small.

\subsection{E. Phase transition in the sub-ohmic spin-boson model}

\begin{figure}
    \centering
    \includegraphics[width=.585\textwidth]{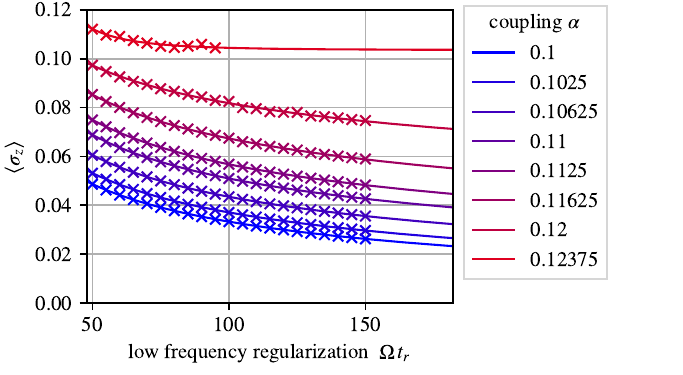}
    \caption{$\sigma_z$ expectation value according to the next-to-leading eigenvector $\rho_2$ in the spectral decomposition as a function of the regularization time $t_r$. The line plots show the fits with the function \eqref{eq:fit}.}
    \label{fig:sz_fits}
\end{figure}

We provide further details on the analysis of the phase transition in the sub-ohmic spin-boson model. For weak coupling this model has a unique steady state that obeys the present $Z_2$ symmetry \cite{Vojta2005Feb,Ren2022Nov,Xu2022Nov,Wang2010May}. However, the so-called Shiba relation predicts an algebraic decay of $\braket{\sigma_z}(t)$ due to the algebraic decay of the bath correlation function \cite{Xu2022Nov}. Hence, the exact spectrum cannot be gapped because this would imply exponential decay. Thus, a rigorous characterization of the phase transition is more involved. For the $\sigma_z$ expectation value we can introduce a continuum approximation of the spectral decomposition \eqref{eq:spectral_decomp}
\begin{equation}
    \braket{\sigma_z}(t)=\sum_{k=1}^{\chi d^2} \eul^{\gamma_k t}\braket{\sigma_z}_k\approx\int_{\gamma\leq 0}\diff\gamma \eul^{\gamma t}J_{\braket{\sigma_z}}(\gamma)+\braket{\sigma_z}_\mathrm{1}.
\end{equation}
We have replaced the sum over eigenvalues by an integral introducing the $\sigma_z$-spectral density $J_{\braket{\sigma_z}}(\gamma)$ from which we have explicitly excluded the unique stationary state contribution. Note that due to the symmetry of the model $\braket{\sigma_z}_\mathrm{1}=0$. Algebraic decay towards this stationary value can be realized when the spectral density behaves as $\gamma^y$ for small $\gamma$ with some exponent $y$. Assuming a smooth $\sigma_z$-spectral density implies that $\braket{\sigma_z}(t)\rightarrow 0$ for large $t$. However, above the phase transition we find a nonzero stationary value if the initial state is nonsymmetric due to spontaneous symmetry breaking. The only way for this to occur is when a second eigenvector contributes significantly to the asymptotic state. This contribution must then be excluded from the continuum approximation
\begin{equation}
    \braket{\sigma_z}(t)\approx\int_{\gamma\leq 0}\diff\gamma \eul^{\gamma t}\tilde J_{\braket{\sigma_z}}(\gamma)+\eul^{\gamma_2 t}\braket{\sigma_z}_{2}.
\end{equation}
In the symmetry broken phase we expect both, $\gamma_2=0$ and $\braket{\sigma_z}_{2}\neq 0$. For the numerics we have to include a low frequency regularization to the bath correlation function. We realize this by enforcing an exponential decay after time $t_r$
\begin{equation}
    \alpha_r(t)=\frac{\alpha(t)}{1+\eul^{-\delta(t_r-t)}}
\end{equation}
($\delta=0.2\Omega$) in order to ensure stability of the algorithm. The true algebraic decay is recovered when $t_r\rightarrow\infty$. Thus, we have to extrapolate to this limit in order to recover the true value of $\braket{\sigma_z}_2$ in the model. In Fig.~\ref{fig:critical_coupling} in the main text we show the low-frequency spectrum of the spin boson model above and below the phase transition and for different $t_r$. We can easily identify the rate $\gamma_2$ and that it vanishes for large $t_r$. The phase transition occurs in the eigenvector $\rho_2$ which changes nonanalytically from $\braket{\sigma_z}_2=0$ to $\braket{\sigma_z}_2\neq 0$. Close to the second order transition we expect power law scaling with $t_r$ (finite size scaling). Hence, we fit our numerical data with
\begin{equation}\label{eq:fit}
    \braket{\sigma_z}_2=\frac{a}{t_r^{b}}+c.
\end{equation}
The fit parameter $c$ should then become nonzero above the critical coupling. As shown in Fig.~\ref{fig:sz_fits} we find very good agreement of our data with the fitted functions, confirming the algebraic behavior. From the predicted asymptotic values of $\sigma_z$ (fit parameter $c$) in Fig.~\ref{fig:critical_coupling} (main text) we can identify the critical coupling. Note that the obtained asymptotic values may only be valid in the vicinity of the transition where algebraic behavior is guaranteed. 

\end{document}